 \documentstyle[twoside,fleqn,espcrc2]{article}
\newcommand{\beq}{\begin{equation}}
\newcommand{\eeq}{\end{equation}}
\newcommand{\bea}{\begin{eqnarray}}
\newcommand{\eea}{\end{eqnarray}}

\newcommand{\tr}{{\rm tr}} 
\newcommand{\sign}{{\rm sign}}

\newcommand{\vev}[1]{\big\langle #1 \big\rangle}
\newcommand{\pars}{\partial^\ast}
\newcommand{\thcb}{\theta_{\pars p}}

\input epsf

\newcommand{\AmS}{{\protect\the\textfont2
  A\kern-.1667em\lower.5ex\hbox{M}\kern-.125emS}}
\title{Bound on the string tension by the excitation probability 
for a vortex}

\author{Tam\'as G. Kov\'acs\address{Instituut-Lorentz for 
                  Theoretical Physics, University of Leiden,\\ 
             P.O. Box 9506, NL-2300 RA Leiden, The Netherlands}
        and 
        E. T. Tomboulis\address{Physics Department, University of 
              California, Los Angeles, CA 90095-1547}
             \thanks{Talk presented by the last author at Lattice 99, 
                     Pisa, Italy.}}
       
\begin{document}

\begin{abstract}
A lower bound on the string tension for large beta in SU(2) 
LGT is derived. The derivation is from first principles and 
bounds the string tension from below by the expectation for 
the excitation of a single `tagged' thick vortex winding around 
the lattice. Thus confinement follows if this expectation remains 
nonvanishing at large beta. Numerical simulations are 
presented to show that this is indeed the case. 
\vspace{1pc}
\end{abstract}

\maketitle

Over the last two years the center vortex picture of 
confinement has undergone  substantial development by  
a series of numerical investigations as well as new 
analytical results (see e.g. \cite{lat} and references therein, 
and contributions to these proceedings). 
Here we report on a new analytical result relating the 
existence of nonzero string tension to the excitation 
probability of a vortex at weak 
coupling. The relation implies that {\it nonvanishing 
expectation for an (arbitrarily) long vortex is a sufficient 
condition for confinement at weak coupling.} 
We then present a measurement of this probability 
by numerical simulations.  

We consider the Wilson loop $W[C] = \vev{\,\tr U[C]\,}$ 
in the $SU(2)$ LGT with plaquette action  
\beq 
A_p(U) = \beta_A\, |\tr U_p| + \beta\: \tr U_p \,,  
\eeq 
and for large $\beta$, $\beta_A$. 
Our result is the following bound on the Wilson loop: 
\beq
W[C]  \leq \exp(\,-\rho(\beta)\,|A|\,) \label{I0}
\eeq 
where $|A|$ is the minimal loop area, and 
\bea  
\rho(\beta)& = &(\mbox{Const})\,\ln [\,1 + 
\vev{\theta_{\pars p}}^{(+)}
\tanh K_0\,] \nonumber\\
    &\approx &(\mbox{Const})\;e^{-4\beta}\,\vev{\thcb}^{(+)}
\;.\label{strtension}
\eea 
with $K_0\equiv {1\over2} \ln \coth (2\beta)$. 

The quantity $\vev{\thcb}^{(+)}$ denotes 
the expectation of an operator creating magnetic flux 
forced to wind completely around the (periodic) lattice. 
It is defined as follows. 

Define for a $3$-cube $c$: 
\beq
\eta_c=\prod_{p\in\partial c}\, \mbox{sign}\,\tr U_p \;.
\eeq
A configuration such that $\eta_c=-1$ represents a Dirac 
monopole of $Z(2)$ flux residing on the cube $c$. The set of cubes on 
which $\eta_c=-1$ must form coclosed sets, i.e. closed sets 
on the dual lattice (Bianchi identity - magnetic current 
conservation). 
\begin{figure}[h]
\begin{minipage}{70mm}
{\ }\hfill\epsfysize=4cm\epsfbox{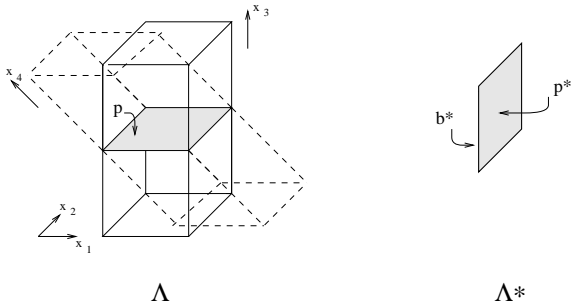}\hfill{\ }\\
\vspace{-1.5cm}\caption{\label{cb} Coboundary of plaquette $p$ consists of 
the set of cubes having the plaquette (shaded) on their 
boundary. On the dual lattice, this corresponds to 
the boundary of the dual plaquette $p^\ast$ (d=4). 
The smallest monopole loop has $\eta_c=-1$ on the cubes of such   
a single plaquette coboundary.}
\end{minipage}
\end{figure}
Thus in $d=4$ (where a cube is dual to a bond), a 
coclosed set of cubes is a closed loop of dual bonds: 
a $Z(2)$ monopole loop. The smallest such set is the coboundary 
of a plaquette (figure \ref{cb}).  

By virtue of the conservation constraint, the presence 
(absence) of a minimal length monopole `loop' 
on the coboundary $\pars p$ of a plaquette $p$  
is characterized by $\thcb=1$ ($\thcb=0$), where   
\beq
\thcb \equiv 
\prod_{c\in\,\pars p}\;{1\over2}[\,1 - \eta_c\,] \,.\label{thcb}
\eeq  
It is the expectation $\vev{\thcb}^{(+)}$ of this quantity 
(\ref{thcb}), for some fixed plaquette $p$, that 
occurs in (\ref{strtension}). The choice of $p$ is 
irrelevant by translational invariance. The `$+$' in 
$\vev{\ }^{(+)}$ signifies that the expectation of $\thcb$, 
is computed with:  
\begin{enumerate}
\item[(i)] plaquette action: $(\beta_A + \beta)\, |\tr U_p|$;
\item[(ii)] $\eta_c=1$ for all $3$-cubes on the lattice 
other than the cubes belonging to $\pars p$ in the numerator, 
and for all $3$-cubes in the denominator (i.e. the partition 
function) of the expectation; 
\item[(iii)] $\sign\, U_{p^\prime} =1$ for every plaquette $p^\prime$ 
on the $2$-dimensional plane $S$ spanning the lattice 
and containing the plaquette $p$ of $\pars p$.  
\end{enumerate} 
It is important to note, and easily seen that, with periodic 
boundary conditions, (i)-(iii) imply that $\vev{\thcb}^{(+)}$ 
depends only on coset $SU(2)/Z(2)\sim SO(3)$ rather than 
$SU(2)$ bond variable configurations.     
\begin{figure}[htb]
\begin{minipage}{7cm}
{\ }\hfill\epsfysize=4cm\epsfbox{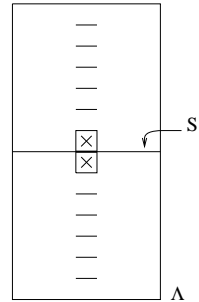}\hfill{\ }\\
\vspace{-1cm}\caption{\label{tv} Tagged vortex winding 
around the lattice.} 
\end{minipage}
\end{figure}

The expectation $\vev{\thcb}^{(+)}$ is now seen to have the 
following physical meaning. (We refer to \cite{KT}, \cite{Y} for a 
physical discussion of vortices and constraints on fluxes on 
the lattice.) Constraint (iii) 
forbids any  net vortex flux from 
crossing the surface $S$. Furthermore, 
monopole excitation is forbidden everywhere by constraint (ii)
except on the coboundary of the plaquette $p$ on $S$, 
where the factor $\thcb$ enforces the presence 
of a monopole loop.  The Dirac sheet in the $SO(3)$ 
configuration, representing the vortex flux attached to this 
monopole loop, is thus forced to wind around the periodic 
lattice in the perpendicular directions.      
$\vev{\thcb}^{(+)}$ is then the excitation probability amplitude 
for a vortex completely winding 
around the lattice in the directions perpendicular 
to the surface $S$ `anchored' by a minimal-length monopole loop, 
encircing $S$ (figure \ref{tv}). This is in fact a version  
of a 't Hooft loop operator \cite{tH}. Shrinking the monopole loop 
to a point, would result in the flux of a complete, `unpunctured' 
vortex trapped inside the periodic lattice.   

Eqs. (\ref{I0})-(\ref{strtension}) then imply nonvanishing 
string tension {\it provided} the expectation $\vev{\thcb}^{(+)}$ 
remains nonvanishing in the large volume limit at (arbitrarily) 
large $\beta$. What decides this is whether the free-energy 
cost of the flux forced to wind around the (periodic) lattice 
remains finite in the large lattice limit. This can happen if  
flux can efficiently spread out in the transverse directions 
to compensate for the cost along the winding directions \cite{KT}, 
\cite{Y}. The monopole loop itself represents  
a local effect of fixed action cost, and serves to `anchor' and 
tag the vortex. This is a very convenient device that leads 
to derivation of (\ref{I0})-(\ref{strtension}). 

Indeed, the derivation relies on a factorization inequality 
for the expectation of $n$ tagged vortices in terms of the 
product of the expectation of $(n-1)$ tagged vortices 
times that of a single tagged vortex.  It further 
uses the $SO(3)\times Z(2)$ formulation of the 
$SU(2)$ LGT \cite{KT}, and subsequent duality transformation 
on the $Z(2)$ part. The somewhat lengthy details will appear 
elsewhere. 

The finite local cost associated with the monopole 
loop site can in fact be explicitly extracted, and then  
$\vev{\thcb}^{(+)}$ can be related to the expectation for a 
single unpunctured vortex winding around the lattice, i.e. 
essentially 't Hooft's magnetic-flux free-energy order 
parameter \cite{tH}.  Now both the latter and 
$\vev{\thcb}^{(+)}$ are nonvanishing 
in the large volume limit already when evaluated in the semiclassical 
approximation for spacetime dimension $d\leq 4$. 
One expects this to persist and in fact be improved in the 
full theory where flux can spread out nonperturbatively. 
In the absence of an analytical proof, we have resorted to 
numerical evaluation of the vortex excitation 
expectation (magnetic flux free-energy).  
\begin{figure}[!htb]
\begin{minipage}{7cm}
{\ }\hfill\epsfxsize=7cm
\epsfysize=6.5cm\epsfbox{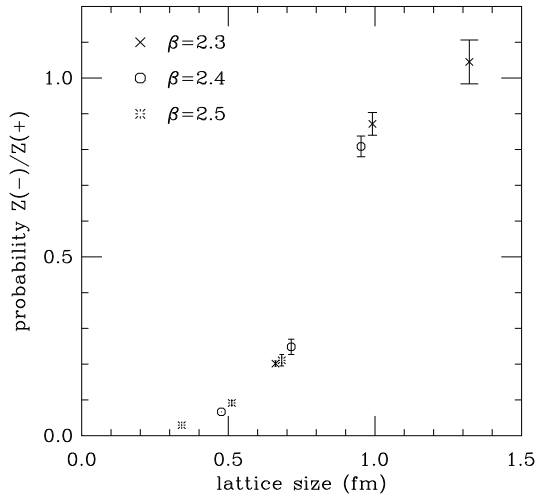}\hfill{\ }\\
\vspace{-1cm}\caption{\label{vfe} Vortex probability (magnetic-flux 
free-energy) vs. lattice size} 
\end{minipage}
\end{figure}
\vspace{-0.3cm}

Measurement was performed by the multihistogram method \cite{FS}.
The method tends to be computationally expensive. It was used 
in \cite{HRR} to compute the free energy of a Z(2) monopole 
pair as a function of the pair's separation.  
The result of our computation is shown in figure \ref{vfe}. 
The lattice spacings are $a=0.119$ fm and $a=0.085$ fm for $\beta=2.4$ 
and $\beta=2.5$, respectively. As expected by physical reasoning, not 
only does the vortex free energy cost remain finite as 
the lattice volume grows, but it tends to 
zero, i.e. the weighted probability for the presence of a vortex 
goes to unity for sufficiently large lattice. This reflects 
the exponential spreading of color-magnetic flux in a confining phase.  

In conclusion, we have seen that nonvanishing expectation for 
an arbitrarily long spread-out vortex is sufficient to ensure 
that the $SU(2)$ LGT remains in a confining phase at  
any large $\beta$. This should be combined with the  
result that the presence of vortices at weak coupling is also 
necessary for confinement: constraining a Wilson loop to be insensitive to 
the presence of thick vortices linking with it leads to nonconfining 
behavior \cite{KT1}.  
Both these statements are clearly demonstrated in the numerical 
simulations. In fact, the measurement of the magnetic-flux free-energy 
presented here is interesting in its own right as it also 
shows that the probability at large $\beta$ for the presence 
in the vacuum of a sufficiently spread-out vortex actually  
tends to one. This is in complete agreement with the results 
of the  closely connected measurements in \cite{HRR}.

\vspace{0.3cm}

This work was supported by FOM, and NSF grant NSF-PHY 9531023.

\end{document}